\documentclass[11pt,letterpaper]{article}
\usepackage[T1]{fontenc}
\usepackage[utf8]{inputenc}
\usepackage[hmargin=.94in,top=.85in,bottom=.88in]{geometry}
\usepackage{newtxtext,newtxmath}
\usepackage{microtype}
\usepackage{graphicx}
\usepackage{booktabs,array,tabularx,enumitem,etoolbox}
\usepackage[font=small,labelfont=bf]{caption}
\usepackage[numbers,sort&compress]{natbib}
\usepackage{xurl}
\usepackage{xcolor}
\usepackage[hidelinks]{hyperref}
\providecommand{\doi}[1]{\href{https://doi.org/#1}{\nolinkurl{doi:#1}}}
\hypersetup{pdftitle={LLM Agents for Financial Knowledge Graphs: A Survey}}
\setlist{nosep,leftmargin=*}
\title{LLM Agents for Financial Knowledge Graphs: A Survey}
\newcommand{\surveyAuthorBlock}{%
  \begin{minipage}{\dimexpr\textwidth-2\tabcolsep\relax}
    \centering
    \normalsize
    Yunbai Zhang\textsuperscript{1,*,\textdagger}\quad
    Heze Ma\textsuperscript{2,*}\quad
    Nan Jiang\textsuperscript{3,*}\quad
    Wenxuan Liu\textsuperscript{1}\\[0.25em]
    Ze Chen\textsuperscript{1}\quad
    Yunyi Gao\textsuperscript{1}\quad
    Winnie Gao\textsuperscript{1}\quad
    Han-Tai Shiao\textsuperscript{1}
    \par\vspace{0.5em}
    \small
    \textsuperscript{1}The Vanguard Group, Malvern, Pennsylvania, USA\\
    \textsuperscript{2}Columbia University, New York, New York, USA\\
    \textsuperscript{3}University of Miami, Coral Gables, Florida, USA
    \par\vspace{0.4em}
    \textsuperscript{1}\{\href{mailto:yunbai_zhang@vanguard.com}{\nolinkurl{yunbai_zhang}},
    \href{mailto:wenxuan_liu@vanguard.com}{\nolinkurl{wenxuan_liu}},
    \href{mailto:ze_chen@vanguard.com}{\nolinkurl{ze_chen}},
    \href{mailto:yunyi_gao@vanguard.com}{\nolinkurl{yunyi_gao}},
    \href{mailto:winnie_gao@vanguard.com}{\nolinkurl{winnie_gao}},
    \href{mailto:han-tai_shiao@vanguard.com}{\nolinkurl{han-tai_shiao}}\}@vanguard.com\\
    \textsuperscript{2}\href{mailto:hm2938@columbia.edu}{\nolinkurl{hm2938@columbia.edu}}\quad
    \textsuperscript{3}\href{mailto:nxj286@miami.edu}{\nolinkurl{nxj286@miami.edu}}
  \end{minipage}%
}
\author{\surveyAuthorBlock}
\hypersetup{pdfauthor={Yunbai Zhang, Heze Ma, Nan Jiang, Wenxuan Liu, Ze Chen, Yunyi Gao, Winnie Gao, Han-Tai Shiao}}
\date{}
\makeatletter
\patchcmd{\@maketitle}{\vskip 2em}{\vskip .5em}{}{}
\patchcmd{\@maketitle}{\vskip 1.5em}{\vskip 1em}{}{}
\pretocmd{\@makefntext}{\hangindent=1.8em\relax\hangafter=1\relax}{}{}
\makeatother
\begin{document}
\maketitle
\begingroup
\renewcommand{\thefootnote}{\fnsymbol{footnote}}
\footnotetext[1]{These authors contributed equally.}
\footnotetext[2]{Corresponding author.}
\endgroup

\begin{abstract}

Financial knowledge graphs (KGs) link information across sources to support risk assessment, investment research, asset recommendations, and forecasting. Large language models (LLMs) help build these graphs by extracting relationships from text and matching entities across records. Yet a workflow may still prescribe which sources to consult and which checks to run. If the chosen sources lack relevant evidence, even accurate extraction can leave a financial question unanswered. LLM agents can adapt the workflow as evidence emerges. They use the graph and earlier findings to decide where to look next, which claims to verify, and whether to update a record for later use. This is the first survey devoted to how LLM agents reshape the construction, maintenance, and use of financial KGs. Our taxonomy covers \emph{Knowledge Graph Construction}, \emph{Temporal Evolution}, \emph{Verification, Reconciliation, and Admission}, \emph{Adaptive Acquisition}, and \emph{Knowledge Graph Usage}. We examine how studies evaluate graph quality, agents' use of evidence, financial task outcomes, and reliability over time. These evaluations show that agents can recover evidence missed in an initial search and revise records as new information arrives. Three questions remain unresolved. How can an agent tell whether it has enough evidence for a financial decision? After a record is corrected, how can the agent identify which earlier conclusions need to change? Which relationships and rules can it reuse in a new graph or domain?
The papers, datasets, and benchmarks covered in this survey are listed at
\mbox{\textcolor[HTML]{0645AD}{\url{https://github.com/YunbaiZhang/Financial-Knowledge-Graph-Survey}}}.

\end{abstract}

\section{Introduction}\label{sec:1}

Financial knowledge graphs (KGs) link information across sources so analysts can trace financial relationships to the evidence behind them. Building and maintaining such a graph requires matching entities across sources, checking which relationships the evidence supports, and keeping those relationships current. Large language models (LLMs) make it easier to extract and match information expressed in different ways~\citep{constructionSurvey,edc}. Even when an LLM handles extraction and matching, the workflow may still prescribe which sources to consult and which checks to run. LLM agents can adapt those choices to the task. What they find in the graph can guide where they search next, which claims they verify, and whether they propose an update~\citep{swissRegistry,finreflectkg}.

Existing surveys examine parts of this process. \emph{LLM-Empowered Knowledge Graph Construction}~\citep{constructionSurvey} covers schema design, extraction, and integration. \emph{Temporal Knowledge Graph Completion}~\citep{temporalCompletion} examines how to infer missing facts in evolving graphs. \emph{Graph-based Agent Memory}~\citep{graphMemory} studies how agents store, retrieve, and update knowledge and experience across tasks. \emph{Integrating Graphs, Large Language Models, and Agents}~\citep{integratingGraphs} compares ways to combine graphs and LLMs for reasoning and retrieval. It examines which approaches suit different tasks and graph types.

In finance, it is particularly important to trace a decision to the evidence available when it was made. \emph{Conceptual Modeling and Semantic Relations in the Construction of Financial Knowledge Graphs}~\citep{conceptualFinance} compares ways to represent financial relationships and identifies continuous updating as a challenge. \emph{A Comprehensive Review of Financial Knowledge Graphs}~\citep{financialKGReview} compares graph types, construction methods, and applications. It also discusses LLM-assisted methods and calls for an audit trail of graph updates.

Building on these perspectives, we organize the literature around the decisions that connect evidence gathering, graph maintenance, and financial analysis. When a graph leaves a question unresolved, how does an agent decide which external sources to consult and how to verify what it finds? When is that evidence sufficient for the current answer, and when does it also justify updating a record that later analyses will use? This is the first survey devoted to how LLM agents reshape the construction, maintenance, and use of financial knowledge graphs. We focus on explicit financial KGs, compare agent-based methods with their non-agentic foundations, and draw on general KG and graph-memory studies where financial evaluations remain limited.

The survey makes three contributions.

\begin{enumerate}

\item \textbf{Agent-centric KG taxonomy.} We provide a five-area taxonomy for comparing how LLM agents reshape the construction, maintenance, and use of financial knowledge graphs.

\item \textbf{Quality \& reliability evaluation.} We examine how studies assess graph quality and what their results establish about the reliability of analyses that use the graph.

\item \textbf{Open challenges.} We identify the judgments that remain difficult for financial KG agents and outline priorities for future research.

\end{enumerate}

The reviewed studies show that agents can use feedback to find missing evidence and revise graph records~\citep{swissRegistry,trackmem}. It remains harder to judge whether the evidence fully supports a financial conclusion and whether that conclusion still holds as the evidence changes. Section~\ref{sec:2} introduces the taxonomy and Section~\ref{sec:3} compares methods within its five areas. Section~\ref{sec:4} examines what current evaluations establish, and Section~\ref{sec:5} discusses the remaining challenges.

\section{Taxonomy Overview}\label{sec:2}

\begin{figure}[!htbp]
\centering
\makebox[\linewidth][c]{\includegraphics[width=1.07\linewidth]{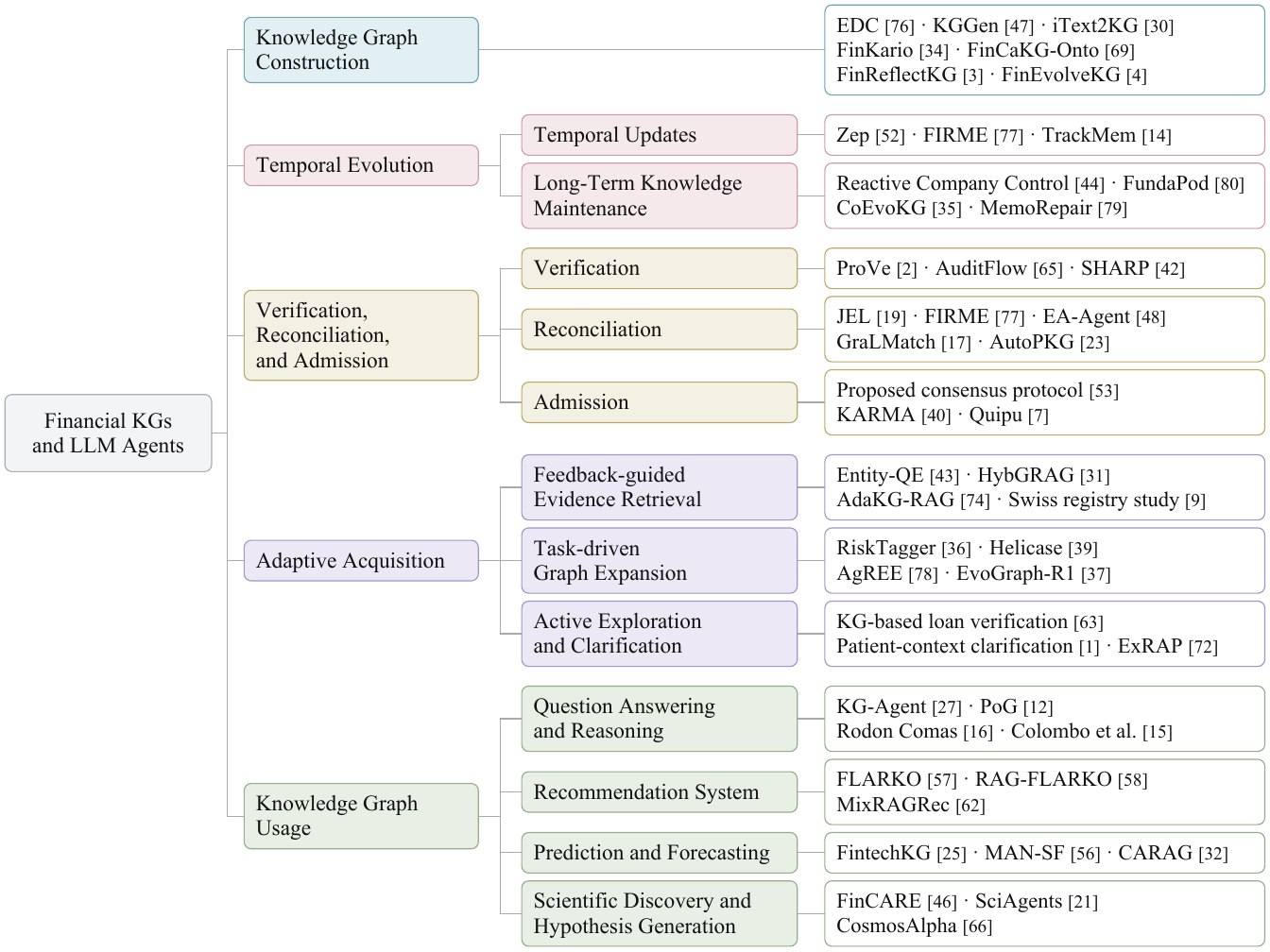}}
\caption{Five-area taxonomy of financial knowledge graphs in the era of LLM agents. The branches follow Section~\ref{sec:3}, with representative studies listed on the right.}
\label{fig:taxonomy}
\end{figure}

We group methods into five areas according to the financial KG problem they address and the result they produce. A single system can span several areas. An agent may search for evidence of a missing relationship and check what the source supports. Deciding whether to add that relationship to the graph is a separate step. Figure~\ref{fig:taxonomy} maps the five areas to their subareas and representative studies.

\textbf{Knowledge Graph Construction} extracts entities and relationships from financial sources and organizes them under a shared schema. Feedback on missing, ambiguous, or unsupported relationships helps agents revise proposed entries or improve instructions for later extractions.

\textbf{Temporal Evolution} keeps financial knowledge current while preserving its history. Agents compare new evidence with earlier records. The evidence may add a dated observation, show that a relationship has changed, or correct an earlier error. This distinction guides which records to revise and which conclusions to revisit.

\textbf{Verification, Reconciliation, and Admission} address three questions. Does the evidence support the claim? Do the records describe the same entity or relationship? Should the proposed change enter the shared graph? Agents can investigate conflicting sources or ambiguous matches. Admission rules determine which proposed changes take effect.

\textbf{Adaptive Acquisition} addresses gaps in the evidence available for a task. Agents can revise a search, consult an external source, follow a new lead, or ask for clarification. The results guide the next search or give the agent a reason to stop.

\textbf{Knowledge Graph Usage} applies graph evidence to financial questions, recommendations, forecasts, and research hypotheses. Agents select and combine relationships relevant to the task. When the evidence supports only part of a conclusion, they revisit their reasoning.

\begin{figure}[!htbp]
\centering
\includegraphics[width=\linewidth]{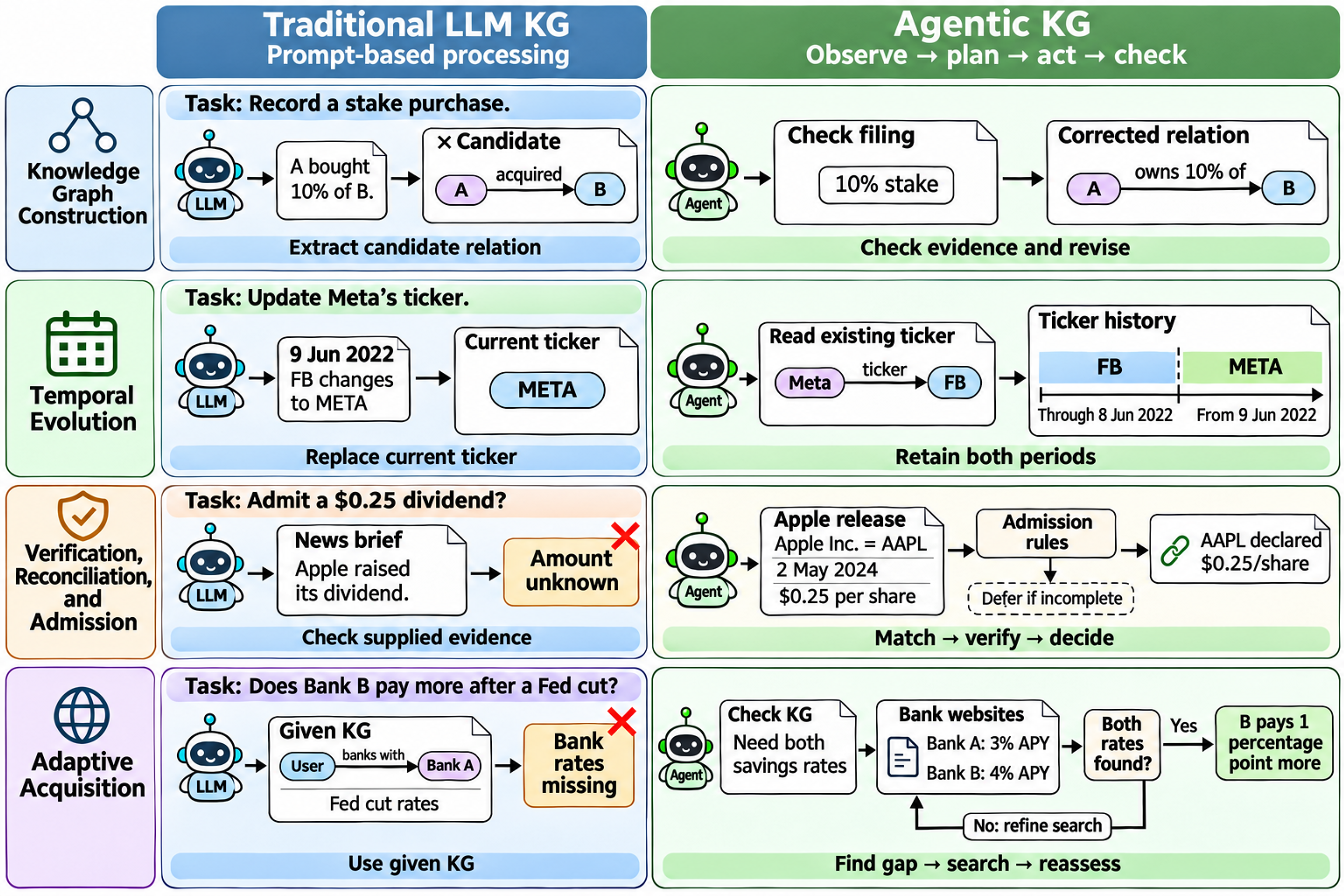}
\caption{Illustrative workflows for traditional LLM KGs and agentic KGs. The agents use the current graph and available evidence to choose their next action. The two banks' savings rates are hypothetical.}
\label{fig:workflows}
\end{figure}

Figure~\ref{fig:workflows} compares how a prescribed workflow and an agent handle the same four financial KG tasks.

These areas meet when a task exposes a gap. New evidence may answer the immediate question. Keeping it for later use requires a separate judgment about what the graph should record and for which period.

\section{Taxonomy in Detail}\label{sec:3}

\subsection{Knowledge Graph Construction}\label{sec:3.1}

Knowledge graph construction organizes source information into entities and relations under a shared schema. EDC~\citep{edc} standardizes relation labels, KGGen~\citep{kggen} removes duplicate entities, and iText2KG~\citep{itext2kg} integrates new information into an existing graph. Financial applications include extracting company events from research reports with FinKario~\citep{finkario} and causal statements from annual filings with FinCaKG-Onto~\citep{fincakg}.

Extraction and normalization must work together when a proposed relation conflicts with existing knowledge, could match several graph concepts, or has weak source support. In REFinD~\citep{refind}, for example, models mistook partial equity ownership for an acquisition. Agentic workflows use feedback to revise candidate triples or the instructions that guide extraction.

FinReflectKG~\citep{finreflectkg} revisits triples extracted from each filing passage. A critic checks entity and relation labels against the financial schema, and a correction model revises or drops flagged triples. Review ends when the critic finds no further issues or the iteration limit is reached. FinReflectKG achieved higher schema compliance than single-pass or fixed two-pass extraction in the reported evaluation. FinEvolveKG~\citep{finevolvekg} uses feedback at a different stage. During development, it uses missed facts, unsupported relations, and type errors to revise analyst-specific extraction instructions. The shared grounding and typing instructions stay fixed during this role-specific adaptation. The revised instructions then guide extraction from new filings.

\subsection{Temporal Evolution}\label{sec:3.2}

Temporal evolution updates time-dependent graph facts while preserving their history. A relationship that has ended can still be valid for an earlier period. An error in the original record, however, requires correction. Temporal Record Linkage~\citep{temporalLinkage} tracks entities as their attributes change, while Time-Aware Probabilistic KGs~\citep{probabilisticTime} distinguish when facts hold from when they are recorded. In finance, TKGQA~\citep{tkgqa} represents changes from merger news through successive graph snapshots.

\subsubsection{Temporal Updates}\label{sec:3.2.1}

Updating temporal knowledge requires determining how incoming evidence relates to existing records. In financial dialogue, a statement may revise a judgment, reinforce a recurring interest, or add a dated observation. Agent memory systems use the graph to help an LLM interpret the statement and identify the affected records.

Zep~\citep{zep} uses an LLM to compare incoming facts with related stored edges. If conflicting facts overlap in time, Zep marks the older edge as valid only until the new fact takes effect. It keeps the old edge so that earlier graph states can still be retrieved.

The evidence for an investor's interests can strengthen or weaken over time. FIRME~\citep{firme} links new observations to a relationship's history using stored aliases and, when needed, LLM matching. Fixed rules adjust the strength of the evidence as observations lapse or recur. A period of inactivity alone does not establish that an interest has ended.

Beyond recurring interests, financial dialogue requires distinguishing revised judgments from new observations. TrackMem~\citep{trackmem} uses an LLM to identify which existing quantity a new target price revises. Its rules mark the previous value as historical and record when it ceased to apply. Dated market observations remain separate. The agent can then retrieve both current and historical judgments.

\begin{figure}[!bp]
\centering
\includegraphics[width=0.92\linewidth]{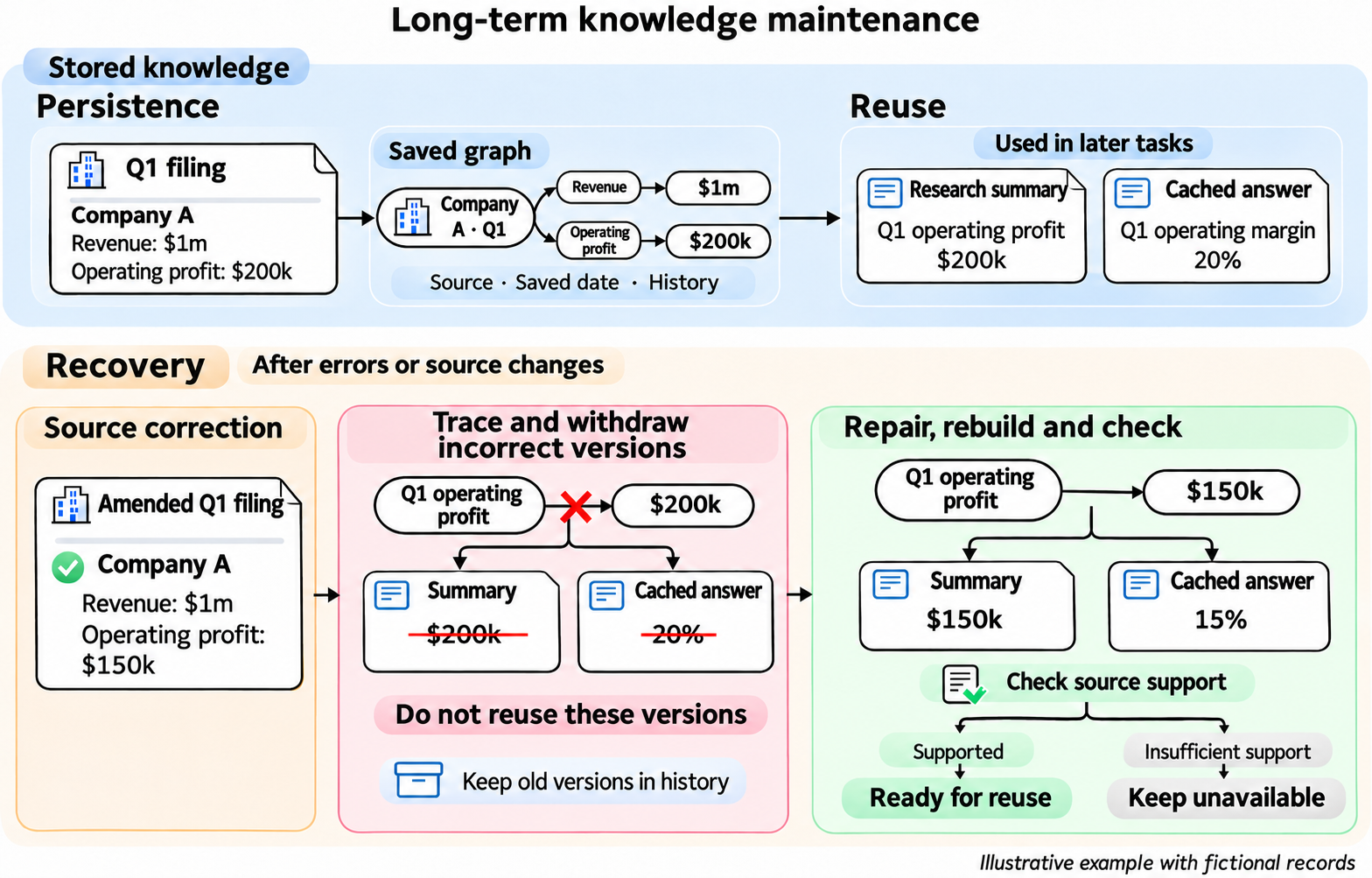}
\caption{Persistence, reuse, and recovery in a financial KG. An amended filing lowers operating profit from \$200,000 to \$150,000. The graph retains the original record in its history, and the corrected profit updates both the research summary and the cached operating margin.}
\label{fig:maintenance}
\end{figure}

\subsubsection{Long-Term Knowledge Maintenance}\label{sec:3.2.2}

Long-term knowledge maintenance preserves accumulated knowledge for later use and propagates source changes to dependent graph relations and outputs. After an update is accepted, the system needs to determine which conclusions remain supported, which outputs require rebuilding, and which historical evidence should remain accessible. Figure~\ref{fig:maintenance} illustrates this process for an amended quarterly filing.

\textbf{Persistence.} Changes in shareholding can invalidate indirect control relationships. Reactive Company Control~\citep{reactiveControl} uses incremental rules to identify affected conclusions and derive them again if another ownership path still supports them. This maintains the control graph without recomputing every relationship. Maintenance in agent workflows also extends to the research outputs built from stored evidence.

\textbf{Reuse.} FundaPod~\citep{fundapod} retains investment memoranda and source evidence, reconstructing their graph connections so its master agent can revisit earlier research. In general QA, CoEvoKG~\citep{coevokg} carries evidence across training rounds. It adds selected evidence from successful searches to the graph used to generate later tasks and compute rewards. Earlier searches thus shape what the agent can learn in subsequent rounds.

\textbf{Recovery.} Reuse creates dependencies that require reassessment after source corrections. MemoRepair~\citep{memorepair} uses recorded dependencies to identify and withdraw affected summaries and cached outputs. It then rebuilds selected replacements from retained evidence. If a repair fails, its dependent outputs remain unavailable. Recovery across memory records depends on which links the system recorded.

\subsection{Verification, Reconciliation, and Admission}\label{sec:3.3}

\begin{figure}[!htbp]
\centering
\includegraphics[width=0.92\linewidth]{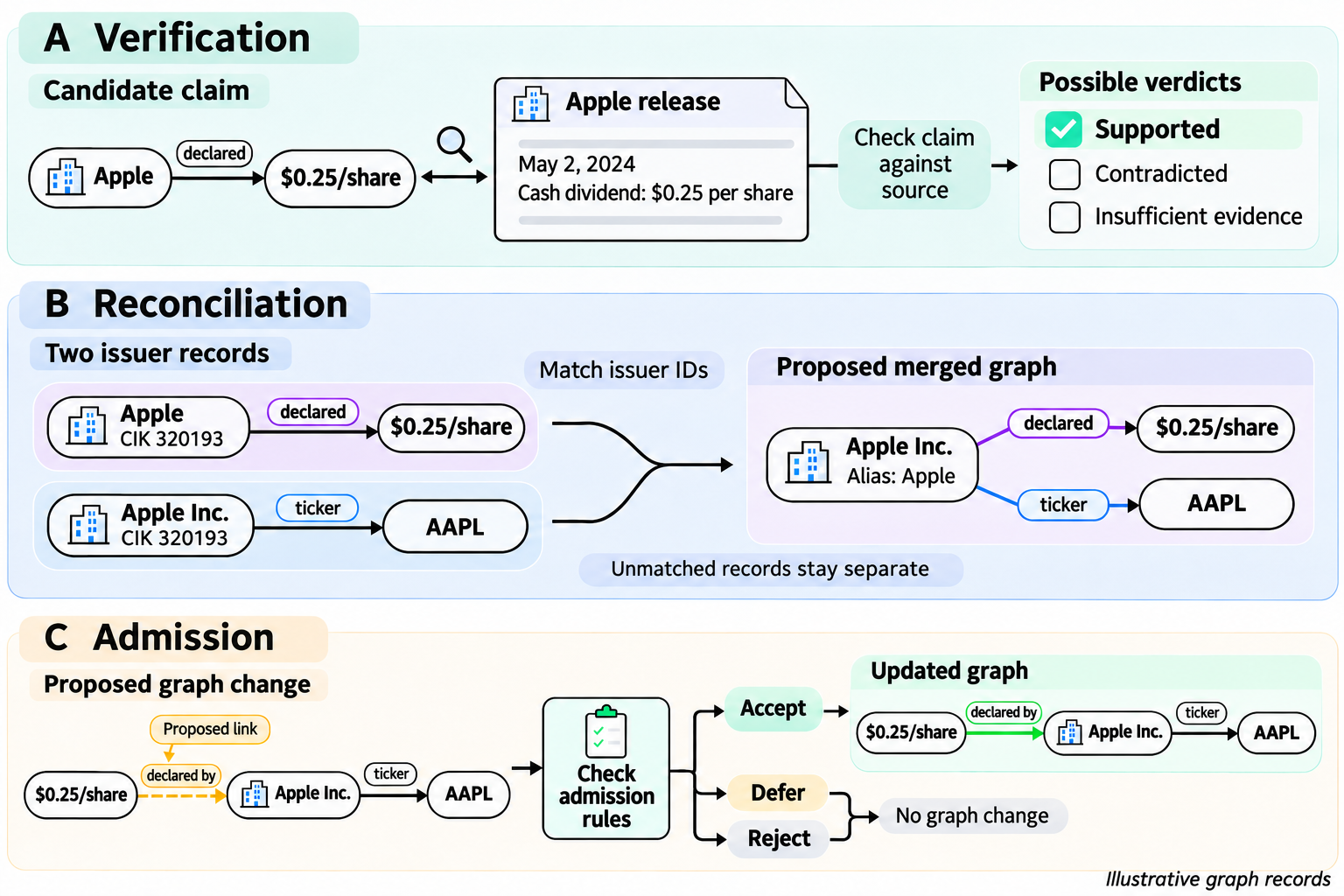}
\caption{Verification, reconciliation, and admission of a dividend record. (A) Apple's \href{https://www.apple.com/newsroom/2024/05/apple-reports-second-quarter-results/}{May 2, 2024 announcement} confirms a cash dividend of \$0.25 per share. (B) A shared issuer ID links two records for Apple. (C) Admission rules determine whether to accept, defer, or reject the proposed dividend relation.}
\label{fig:admission}
\end{figure}

Verification checks the evidence for a claim. Reconciliation determines whether records describe the same entity or relationship. Admission decides which changes to accept into the graph. Verification and reconciliation can inform admission without directly changing the graph. Figure~\ref{fig:admission} illustrates these decisions using Apple's dividend announcement. Section~\ref{sec:3.2} covers temporal updates and maintenance of the resulting records and dependencies.

\subsubsection{Verification}\label{sec:3.3.1}

Verification examines whether a source is reliable, whether its text supports the claim, and whether the claim satisfies relevant constraints. These checks can give different results. A total copied accurately from a financial filing, for example, can still disagree with its component values. ProVe~\citep{prove} selects passages and uses a language model to assess whether the cited text supports a triple. SPARQL-based fund checks~\citep{fundChecks} formalize investment constraints over financial RDF data.

When evidence is missing or checks disagree, the system must decide what to investigate next. AuditFlow~\citep{auditflow} links a US-GAAP taxonomy graph with XBRL filing facts. Two LLM auditors use tools to investigate accounting constraints and filing evidence separately. If their reports disagree, a senior auditor directs further checks.

For claims requiring both graph and textual evidence, SHARP~\citep{sharp} plans checks using relation schemas and prior verification trajectories. Each result guides the choice of the next tool or the decision to stop. On general KGs, SHARP improved triple verification over retrieval baselines and produced a trace of its investigation.

\subsubsection{Reconciliation}\label{sec:3.3.2}

Reconciliation determines whether different representations refer to the same entity or express equivalent relations. Financial news reports may provide few details to distinguish firms with similar names. JEL~\citep{jel} combines character-pattern matching with contextual neural embeddings to link these mentions to company entities in a financial KG. FIRME~\citep{firme} stores aliases learned from earlier matches in an entity registry and reuses them to reconcile incoming name variants with canonical entities.

For ambiguous entities, a central challenge is deciding which attributes or relationships distinguish plausible matches. EA-Agent~\citep{eaAgent} uses entity statistics and similarities among candidates to select attributes and relations. When the candidates are difficult to distinguish, it uses a reflection step to guide the selection. An LLM then matches entities using the selected triples. This makes evidence selection explicit and improves alignment on general benchmarks.

Matching also affects how records are consolidated. GraLMatch~\citep{gralmatch} removes suspicious links in company and security matching graphs to limit incorrect grouping. AutoPKG~\citep{autopkg} compares proposed product records with canonical nodes in the graph. It uses the comparison to decide which records to merge and which to keep separate. TypeSafe's Jev entity-alignment cookbook~\citep{jev} illustrates pairwise scoring on music-catalog records to recommend a merge, manual review, or no link. These operations establish which records correspond. Verification addresses the claims attached to them.

\subsubsection{Admission}\label{sec:3.3.3}

Admission governs the transition from a candidate to reusable graph knowledge. A federated financial KG needs a way for participating institutions to approve proposed changes to their shared graph. A proposed financial-KG consensus protocol~\citep{consensus} separates proposers from acceptors and distributes changes after majority approval. Admission also requires criteria for accepting a proposed change, rejecting it, or sending it for review.

In biomedical KGs, KARMA~\citep{karma} uses assessments of each candidate to guide that decision. A conflict can lead to rejection or expert review. Combined quality scores determine whether a candidate meets the threshold for inclusion.

Quipu~\citep{quipu} complements candidate assessment with policy enforcement at write time. It checks the proposed graph state against the rules. If it refuses a change, it returns the reason so the agent can revise its proposal and try again. In controlled experiments, writers used this feedback to correct violations of label and vocabulary rules. These checks enforce the declared policies. Assessing factual support requires a separate verification step.

\subsection{Adaptive Acquisition}\label{sec:3.4}

When a KG leaves part of a task unresolved, adaptive acquisition determines what information is missing and how to obtain it. The gap may involve overlooked evidence, an unsupported relation, or an unobserved condition. After each round, the agent judges whether the evidence resolves the gap. If the gap remains, it can revise a search, investigate another branch, or ask for clarification. Other areas of the taxonomy address how to represent, check, maintain, and use the evidence acquired.

\begin{figure}[!htbp]
\centering
\includegraphics[width=0.9\linewidth]{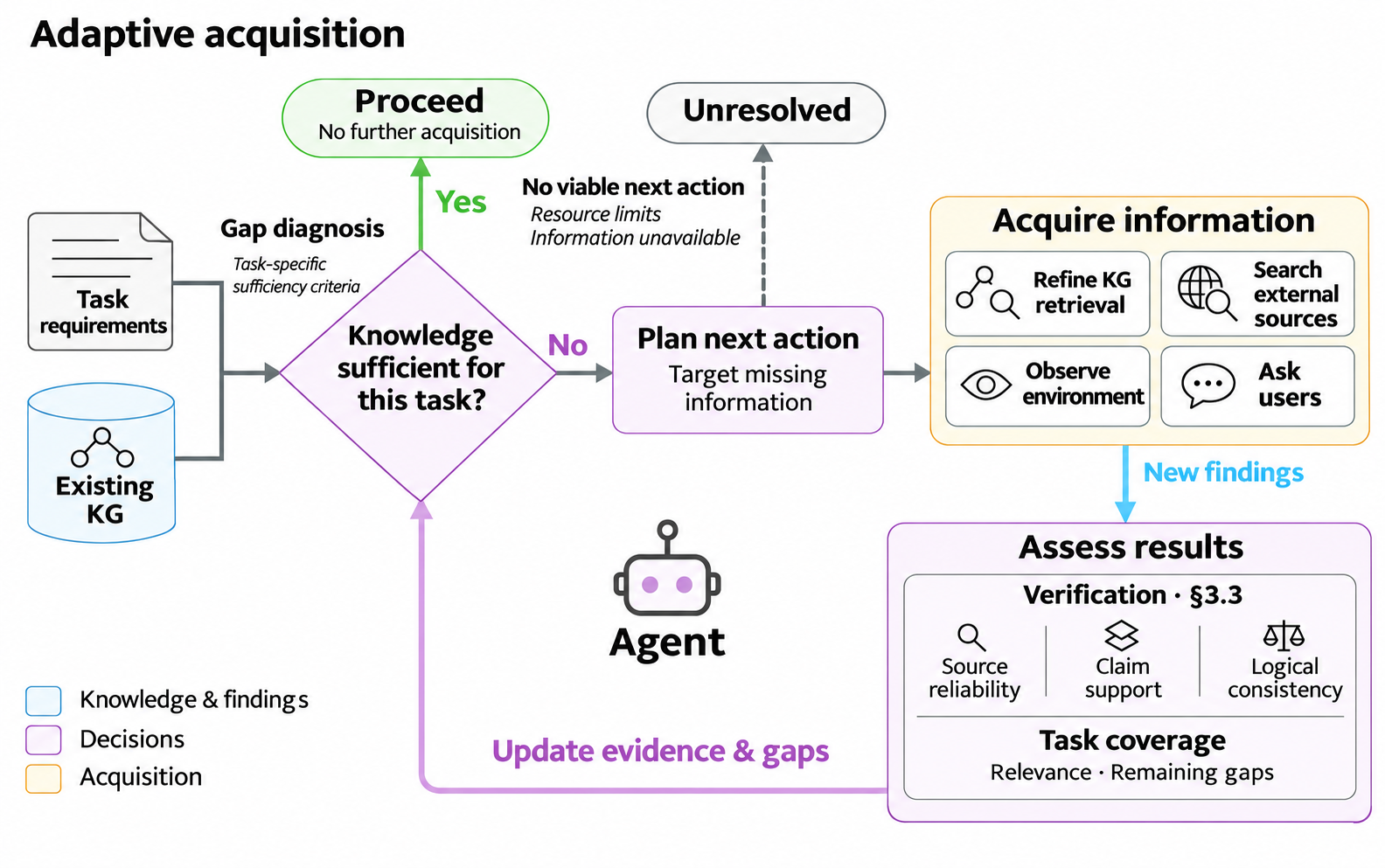}
\caption{\textbf{Adaptive acquisition.} Knowledge gaps guide information acquisition and reassessment. The loop ends with sufficient support or an unresolved task when no viable action remains. In the savings-rate example in Figure~\ref{fig:workflows}, a missing bank quote guides the next search.}
\label{fig:acquisition}
\end{figure}

\subsubsection{Feedback-guided Evidence Retrieval}\label{sec:3.4.1}

A search can miss relevant information even when the source contains it. In financial research, queries may retrieve familiar entities while missing concepts or relationships central to the question. Entity-QE~\citep{entityQE} expands queries with entity names drawn from human relevance judgments, improving document retrieval for query-specific financial KGs. For automated acquisition, the corresponding challenge is generating useful feedback from retrieval results.

KGAgent4CTI~\citep{kgagentCTI} illustrates this mechanism in threat intelligence. Its technique-identification agent assesses whether retrieved MITRE ATT\&CK descriptions provide sufficient evidence. If support is incomplete, it revises the keywords and repeats hybrid retrieval before assigning a technique. Its reported evaluation shows fewer missed and incorrect identifications than single-pass retrieval.

Other methods use feedback to change different parts of a search. HybGRAG~\citep{hybgrag} uses critic feedback to revise the choice of entities, relations, or retrieval modules. AdaKG-RAG~\citep{adaKGRAG} turns missing supporting evidence into follow-up queries. In commercial-registry analysis~\citep{swissRegistry}, failed or empty graph queries similarly prompt revised search terms or a switch to broader text retrieval.

\subsubsection{Task-driven Graph Expansion}\label{sec:3.4.2}

Task-driven expansion addresses entities or relations that the current graph does not represent or adequately support. In financial investigations, rule-based transaction tracing identifies candidate paths. The investigator must then decide which branches to pursue.

RiskTagger~\citep{risktagger} expands an on-chain transaction graph using transaction evidence and case context. At each step, its LLM uses the evidence to recommend whether to keep tracing the funds, while a controller selects the next accounts to examine. This directs expansion toward case-related paths while limiting irrelevant branches.

When several documents describe parts of a relationship, the agent must also decide where to search next. Helicase~\citep{helicase} uses uncertain supply-chain facts and prior search outcomes to prioritize further investigation. Signals that a search would repeat earlier work help direct attention to unresolved relationships.

External retrieval can also supply material for graph expansion. AgREE~\citep{agree} revises searches for emerging entities, while EvoGraph-R1~\citep{evoGraph} incorporates retrieved material into an evolving working graph. Generate-on-Graph~\citep{generateOnGraph} proposes triples from accumulated observations and model knowledge. Each proposed triple still needs supporting evidence.

\subsubsection{Active Exploration and Clarification}\label{sec:3.4.3}

Some information is available only through interaction with a person or the environment. In loan screening, an applicant may repeat supplied personal details without knowing related facts. The system must choose a follow-up question that can elicit new evidence.

A KG-based loan verification system~\citep{loanVerification} uses hierarchical reinforcement learning to select a personal-information node and then a related fact to ask about. Each answer updates the dialogue state in the graph and guides the next question or the decision to stop. In tests with simulated users, this pre-LLM agent achieved higher recognition accuracy than rule-based policies.

For free-form queries, the patient-context clarification framework~\citep{patientClarification} uses an LLM to interpret queries and phrase follow-up questions. A deterministic controller uses information gain to select questions that distinguish graph-supported hypotheses. Each answer then rules out incompatible hypotheses. This acquires missing context before the downstream model responds.

Acquisition can also require actions that produce new observations. ExRAP~\citep{exrap} combines LLM-estimated task value with estimated information value to select environmental actions. The resulting observations update graph memory and guide subsequent choices, improving task success and efficiency in simulated embodied environments.

\subsection{Knowledge Graph Usage}\label{sec:3.5}

KG-based systems can answer financial questions, make recommendations, forecast outcomes, and develop research hypotheses. Each task requires a different combination of evidence. This section examines how agents select and combine graph evidence to meet those requirements.

\subsubsection{Question Answering and Reasoning}\label{sec:3.5.1}

Question answering uses KG facts and relation paths to satisfy a question's conditions. In finance, one answer may require company relationships, reported figures, and news evidence. RoG~\citep{rog} retrieves graph paths from generated relation plans. A retrieved path may satisfy only part of the question. Answering the rest requires deciding which results to combine.

General KG agents use intermediate results to choose their next operation. KG-Agent~\citep{kgAgent} selects executable graph tools from current results and execution history, while PoG~\citep{pog} revisits candidate entities when subgoals remain unresolved. A financial QA agent~\citep{financialQA} combines graph queries, lookups of company fundamentals, and news searches. Results from one source determine the inputs to the next tool. The system checks numerical claims against tool outputs and cites the retrieved passages so readers can trace the evidence.

Financial reasoning also requires explaining derived conclusions. A banking-supervision system~\citep{bankingProof} explains outcomes such as indirect company control through symbolic rules. It identifies recurring patterns in the reasoning and fills explanation templates with facts from the proof. The resulting explanation shows how the conclusion follows from the graph.

\subsubsection{Recommendation System}\label{sec:3.5.2}

Recommendation systems use KG relationships to match candidates to a user's preferences and constraints. In finance, the same market information can support different recommendations when investors' trading histories differ. KGAT~\citep{kgat} aggregates user--item--attribute relations for ranking, while financial momentum models~\citep{financialMomentum} use company relations and market features to rank stocks.

Personalizing these recommendations requires linking market evidence to the investor's own trading history. FLARKO~\citep{flarko} supplies an LLM with personal transaction and market KGs. It uses Kahneman--Tversky Optimization to align recommendations with training labels derived from observed purchases and asset returns. Supplying both graphs in full increases the amount the model must read and leaves it to identify the relevant evidence.

RAG-FLARKO~\citep{ragFlarko} addresses this limitation through selective retrieval. An LLM first selects relevant transaction entities, and SPARQL templates retrieve their subgraph. The resulting subgraph guides the choice of market entities. The stages follow a fixed order, but the market evidence selected depends on the investor's trading history. On FAR-Trans, the retrieval variants achieved a higher hit rate than full-graph prompting for assets that users later bought and that earned positive returns.

Beyond entity selection, MixRAGRec~\citep{mixRagRec} routes recommendation queries to specialist retrievers that search the graph at different levels of detail.

\subsubsection{Prediction and Forecasting}\label{sec:3.5.3}

Prediction and forecasting use graph relations and historical observations to estimate future outcomes. In finance, revenue growth can coincide with declining profitability, and optimistic guidance may be at odds with a firm's past results. A model built on FintechKG~\citep{fintechkg} combines relational, temporal, and financial text features to predict whether revenue will rise or fall. MAN-SF~\citep{mansf} propagates signals from historical prices and social media text through company relations to predict the direction of stock prices.

Earnings-call forecasting also requires deciding how current disclosures should affect a prediction in light of company history and peer evidence. CARAG~\citep{carag} assigns reported results, guidance, and risks to specialist agents that query its temporal KG. The agents compare current performance with historical and sector benchmarks and check whether earlier guidance matched actual outcomes. A synthesizer combines their analyses to predict the direction of returns after the earnings announcement. In the earnings-call evaluation, the full system outperformed the baselines on this task.

\subsubsection{Scientific Discovery and Hypothesis Generation}\label{sec:3.5.4}

Scientific discovery uses graph connections to propose explanations and research hypotheses. In finance, a relationship described in a filing can suggest a causal mechanism without establishing it. FinCARE~\citep{fincare} uses financial KG relations to constrain candidate causal structures and combines statistical discovery with LLM proposals for missing edges. It improves causal-structure recovery on synthetic financial data.

Developing candidate connections into testable research proposals requires deciding which mechanisms to pursue and how to refine them. In materials research, SciAgents~\citep{sciagents} retrieves graph paths, assigns scientist agents to develop proposals, and uses a critic to identify weaknesses and suggest improvements. Its manager selects participants from the evolving discussion, while literature tools inform novelty and feasibility assessments. The resulting proposals specify mechanisms and research directions that remain subject to empirical testing.

CosmosAlpha~\citep{cosmosAlpha} adds quantitative feedback in financial factor discovery. Its research graph connects tested formulas through generation history, structural similarity, and exposure correlation. The system uses evaluation results and graph relationships to choose which formulas the LLM should vary next. Once those variants are evaluated, their results guide the next round of factor proposals.

\section{Evaluation Across the Taxonomy}\label{sec:4}

\begin{figure}[!htbp]
\centering
\makebox[\linewidth][c]{\includegraphics[width=1.07\linewidth]{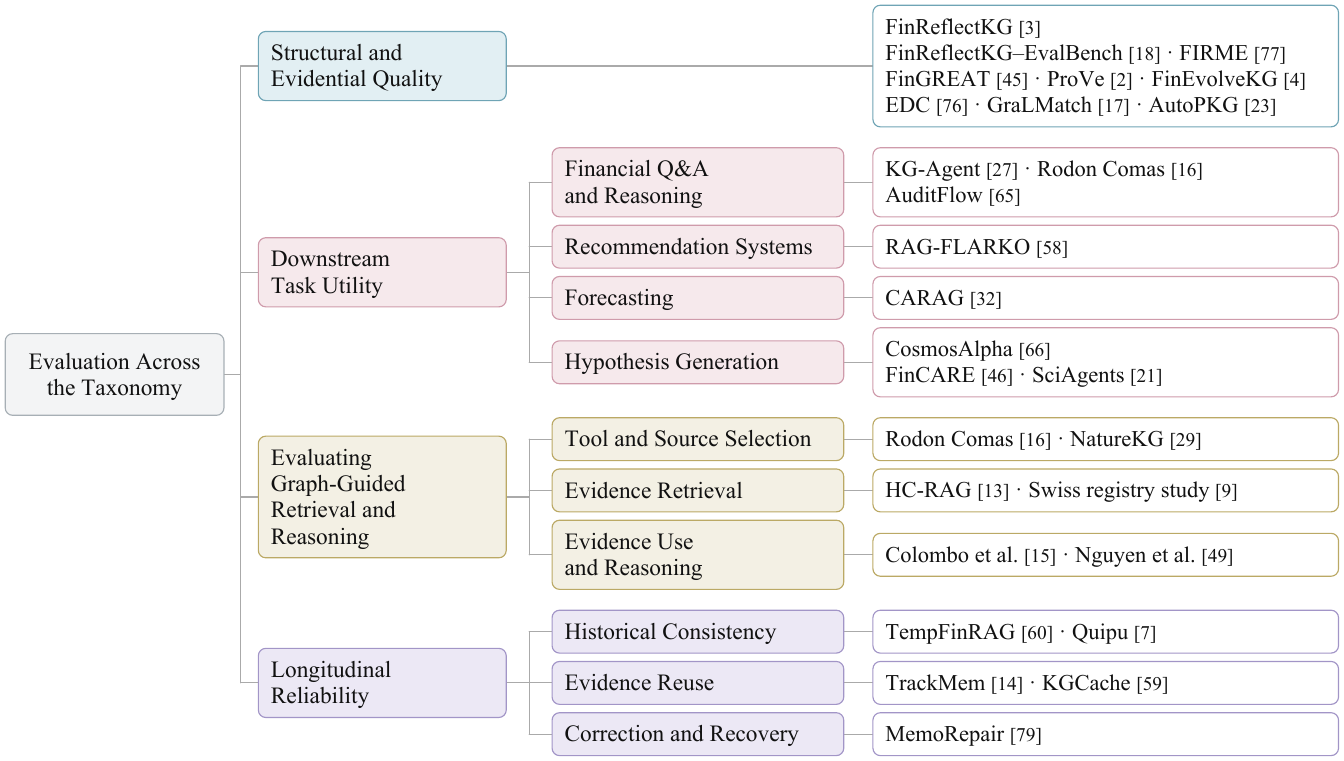}}
\caption{Evaluation of financial knowledge graphs. The branches follow the headings in Section~\ref{sec:4} and list the studies discussed under each heading.}
\label{fig:evaluation}
\end{figure}

Figure~\ref{fig:evaluation} shows how the studies assess the graph itself, the tasks it supports, the evidence an agent uses, and the reliability of its results over time. In Figure~\ref{fig:maintenance}, a historical answer should use the filing available at the time, and the current cached margin should reflect the revised profit.

\subsection{Structural and Evidential Quality}\label{sec:4.1}

Evaluation first asks how faithfully a financial KG represents its sources and which relevant facts it misses. FinReflectKG~\citep{finreflectkg} checks entity types and relation labels against its declared schema. FinReflectKG--EvalBench~\citep{finreflectEval} asks LLM judges to compare each extracted triple with its source passage and rate its support, specificity, and relevance. The benchmark also checks which important facts from each passage are missing from the graph. In this comparison, reflection improves coverage over single-pass extraction. The gain in coverage does not extend to faithfulness to the source.

Checking source support involves both locating evidence and interpreting it. FIRME~\citep{firme} checks whether the quotation attached to an extracted relationship appears in the original text. This establishes traceability, while assessing the relationship requires examining what the passage says. FinGREAT~\citep{fingreat} compares accounting relations predicted from financial passages with the set of relations experts consider acceptable for each passage. ProVe~\citep{prove} checks whether the cited text supports or refutes a claim, or provides too little information to decide. Establishing whether the claim is true independently of its source requires other evidence.

Graph quality also depends on whether the graph contains the facts and connections its users need. FinEvolveKG~\citep{finevolvekg} compares graphs extracted from passages in held-out companies' disclosures with LLM-generated references. It scores general extraction quality separately from coverage of the facts and paths relevant to each predefined analyst role. Experts also review selected outputs. Together, these assessments show how well role-specific extraction captures what an analyst needs while preserving overall quality. Assessing the effect on financial decisions requires a downstream evaluation.

Normalization raises two questions. Do standardized relations preserve the meaning of the source text? Do records grouped together refer to the same entity? EDC~\citep{edc} asks human reviewers whether canonicalized triples remain faithful to the source text. For company and security records, GraLMatch~\citep{gralmatch} compares predicted matches with reference groups. The comparison includes matches implied by grouping linked records, since one incorrect link can cause several records to be assigned the wrong identity. The study repeats the evaluation after cleaning up the groups to see whether fewer records are matched incorrectly.

Evaluation must also address the changes an agent proposes to a financial KG. In product KGs, AutoPKG~\citep{autopkg} compares proposed edits with the consensus of several LLM judges. Human annotations serve as a check on that consensus. The study separately compares extracted product values with their source listings. This distinguishes choosing an appropriate graph edit from extracting its content correctly. Counts of additions, removals, or merges show how much the agent changed. Assessing the resulting graph requires checking whether those edits were correct.

\subsection{Downstream Task Utility}\label{sec:4.2}

Downstream evaluation asks whether using a financial KG improves performance on tasks such as answering questions, recommending assets, forecasting returns, and generating hypotheses.

\subsubsection{Financial Q\&A and Reasoning}\label{sec:4.2.1}

When a question has several valid answers, a reference set can show which entities an agent missed or included in error~\citep{kgAgent}. For supply-chain questions, Rodon Comas's agent~\citep{financialQA} follows supplier and customer links in the KG and combines those relationships with financial data. The evaluation checks the figures in the answer against the values returned by the agent's tools and traces citations to retrieved passages. For filing audits, AuditFlow~\citep{auditflow} links reported facts to the accounting rules that govern them. The study checks whether its agents and a rule-based checker recover the reported and expected values in a reference audit.

\subsubsection{Recommendation Systems}\label{sec:4.2.2}

Historical transaction records allow researchers to compare recommended assets with those an investor later purchased. In RAG-FLARKO~\citep{ragFlarko}, the model selects relevant trades from an investor's transaction KG. Those trades guide its choice of asset price histories to retrieve from a market KG. The study compares the resulting recommendations with those produced when the two graphs are searched independently. It checks both whether investors later bought the recommended assets and whether those assets earned positive returns.

\subsubsection{Forecasting}\label{sec:4.2.3}

Financial forecast evaluation compares predictions with later outcomes. Each prediction must use only information available at the time. CARAG's specialist agents~\citep{carag} query a temporal KG to compare a firm's latest earnings disclosures with its own past results and those of its peers. To evaluate predictions of return direction, the study compares the full system with two alternatives. One removes the KG, and the other gives a single agent access to the same graph. A separate trading simulation assesses the returns and risks of acting on the forecasts.

\subsubsection{Hypothesis Generation}\label{sec:4.2.4}

An agent can propose a new signal for predicting returns or suggest that one financial variable affects another. CosmosAlpha~\citep{cosmosAlpha} uses a graph of earlier formulas to guide an LLM in proposing new signals. It then tests whether those signals, used together, predict returns in a later period. For causal hypotheses, FinCARE~\citep{fincare} combines relationships from a financial KG with links proposed by an LLM. The study checks whether the system recovers the causal links built into simulated company data. The same KG helped shape that simulation. For hypotheses still awaiting an empirical test, SciAgents~\citep{sciagents} uses agent critique and literature searches to assess novelty and feasibility in materials research.

\subsection{Evaluating Graph-Guided Retrieval and Reasoning}\label{sec:4.3}

Evaluating how an agent uses a financial KG requires tracing its choice of sources, the evidence it retrieves, and the reasoning that connects that evidence to a conclusion.

\subsubsection{Tool and Source Selection}\label{sec:4.3.1}

Did the agent choose a suitable source and a tool that could access it? A tool call can succeed even when the source cannot provide the evidence the task requires. In Rodon Comas's financial Q\&A system~\citep{financialQA}, graph queries retrieve supplier relationships and financial data tools retrieve reported figures. Call logs reveal when the agent chose the wrong tool or skipped a necessary lookup. NatureKG~\citep{naturekg} checks the query itself by comparing its relationships, filters, and requested fields with those in a reference query.

\subsubsection{Evidence Retrieval}\label{sec:4.3.2}

Even a search of the right source can miss a fact needed to answer the question. Retrieval evaluation asks whether the evidence returned is relevant and sufficient for the task. HC-RAG~\citep{hcRAG} checks whether a search over financial filings locates the relevant report and the section or table containing the requested figure. For questions spanning several filings, it checks whether the search retrieves evidence from each one. When an agent revises a search, the question is what the next lookup adds. In their study of the Swiss commercial registry, Capozzi and Helbing~\citep{swissRegistry} check whether a later lookup retrieves evidence from the reference answer that the first lookup missed. Matches can include company identifiers and dates.

\subsubsection{Evidence Use and Reasoning}\label{sec:4.3.3}

Evaluating evidence use means checking whether the agent's stated reasoning connects the retrieved facts to its conclusion. An answer can be correct even when the explanation skips a necessary step. Colombo et al.~\citep{bankingProof} compare model-written explanations of corporate control with proofs derived from a financial KG. When an explanation leaves out a detail, it is often an ownership percentage used to establish control. Nguyen et al.~\citep{reasoningPaths} match each stated reasoning step to a graph relation. They then check whether the steps form a connected path to the answer.

\subsection{Longitudinal Reliability}\label{sec:4.4}

Longitudinal evaluation asks whether a financial KG remains reliable as new disclosures arrive, agents reuse earlier evidence, and source records are corrected.

\subsubsection{Historical Consistency}\label{sec:4.4.1}

Can the system reproduce an earlier answer or decision using the information available at the time? A filing about an earlier quarter may have been published after the date of the question. TempFinRAG~\citep{tempFinRAG} asks the same question with dates before and after an SEC filing becomes available. It checks whether later disclosures enter the retrieved evidence or the answer. It also checks whether the system withholds an answer when the required filing is still unpublished. Quipu~\citep{quipu} applies a related test to graph updates. It rechecks previously accepted changes against the data and rules in force at the time and compares the result with the recorded decision.

\subsubsection{Evidence Reuse}\label{sec:4.4.2}

Reuse evaluation asks how much evidence a new task can draw from earlier tasks. It also asks whether the agent uses that evidence correctly and how much repeated search it avoids. In the benchmark used to evaluate TrackMem~\citep{trackmem}, each question is annotated with the earlier statements needed to answer it. The study checks which statements reach the reader, whether its answer is correct, and how many input tokens it uses compared with rereading the full conversation. KGCache~\citep{kgcache} measures repeated search more directly on a general KG. It counts lookups answered from previously retrieved graph neighborhoods and compares retrieval time with and without caching. When the system reuses earlier evidence for a similar question, KGCache also checks the new answer against the benchmark reference.

\subsubsection{Correction and Recovery}\label{sec:4.4.3}

After a correction, does the old value still affect later work? Can the agent produce a valid result from the revised evidence? If a company restates operating income, a financial KG evaluation could revisit a margin based on the original figure. It could then compare that margin with one calculated from the amended filing. MemoRepair~\citep{memorepair} tests this principle in general agent memory. After correcting a source record, it follows recorded dependencies to identify outputs that relied on the old information. It checks which old outputs remain accessible, whether later actions still use them, and how many affected outputs have validated replacements. To measure how much useful information repair restores, the study compares it with simply withdrawing every affected output.

\section{Remaining Challenges and Future Directions}\label{sec:5}

Judging whether a financial KG supports a conclusion remains difficult, even as agents improve at finding and updating the underlying facts. An agent can verify the relationships it finds and still be unsure whether missing evidence would change the answer. After a restatement, it may correct the figure in the graph yet overlook analyses that still rely on the old value. When another graph defines relationships differently, an agent must decide which rules still apply. Over repeated tasks, agents must also judge when reusing evidence preserves answer quality and saves enough work to justify maintaining the graph.

\subsection{Evidence Sufficiency}\label{sec:5.1}

Even when every link an agent finds in a financial KG is correct, it can still be hard to tell whether those links fully answer the question. The agent must decide whether to keep searching based on what it has seen so far.

In a UK case in UBO-Bench~\citep{uboBench}, every ownership chain led to a named shareholder. This satisfied the agent's stopping rule, yet a separate filing said that a person with significant control remained unidentified. Agents flagged this gap more often when explicitly instructed to keep such cases open.

In ABF-Bench~\citep[Sec.~7.5]{abfBench}, the tracing agent kept its own list of leads. In some runs, it closed every lead and submitted its findings with queries to spare. An important branch could still go unexplored if it never appeared on the list. In the incident studied, unexplored branches contained about a quarter of the links the researchers had identified.

Future work should help agents identify which missing relationships could change an answer. They could then pursue those leads and explain what remains unknown when the records fall short.

\subsection{Error Recovery}\label{sec:5.2}

Error recovery in a financial KG means revisiting earlier work after a fact is corrected so that later answers reflect the revised evidence. Agents can trace recorded links from an old value to conclusions that used it. The difficulty is finding every affected conclusion and deciding which ones must change. When an agent uses one conclusion to reach another, it may leave no record of that step. Even if it finds the later conclusion, the evidence it gathered for the original decision may be too limited to tell whether that conclusion still holds.

In Bank-World~\citep[Sec.~5.2.4]{bankWorld}, an agent reported the right decision but gave staff instructions that went beyond the bank's approval. The instructions to loan staff still permitted a larger fee waiver than the bank had approved. In a separate loan review, agents received an amendment but sometimes missed required updates to records or instructions.

In Retinraj's simulations~\citep[Secs.~5.2--5.3]{retinraj}, checking only the direct uses of a changed fact often left the plan incorrect. A later decision could depend on an earlier choice without referring to the fact that shaped it. Both models repaired the plan more reliably when they also followed links between decisions. If the graph omitted such a link, the review could still miss a decision that needed to change.

Progress would mean that a financial KG agent can find every conclusion that relied on a corrected fact. It should revise those that no longer hold and preserve those the revised evidence still supports. Its subsequent answers should consistently reflect the correction.

\subsection{Transferability across Graphs and Domains}\label{sec:5.3}

Transferability is the ability to reuse what an agent has learned from one graph or dataset when it takes on a related task in another domain. In finance, this can reduce the work of defining relationships, writing rules, and labeling data for each new application. The difficult part is deciding what carries over. Two datasets may refer to the same company but record different relationships. An agent must determine which concepts correspond, what evidence supports each relationship, and where a rule from the original graph would lead to a wrong conclusion.

A shared graph schema did not tell FinEvolveKG~\citep[Secs.~6.4--6.5]{finevolvekg} which instructions would help in a new sector. Across U.S. filings, the extractor reused rules for checking whether a passage supported a link. Extra guidance on which links to prioritize for portfolio analysis improved scores on technology filings, but the shared rules alone scored higher in every new sector tested. The scores reveal the better choice in hindsight, but how an agent could make that choice in a new sector remains open.

FGMPL~\citep[Sec.~5.3]{fgmpl} transferred a graph model across markets once researchers had identified the shared relationships. They built a U.S. graph using familiar investment and supply relationships. The researchers kept the Chinese graph encoder fixed. Classification improved with more labeled U.S. examples. How an agent could find the shared relationships or recognize when no match exists remains open. Portisch et al.~\citep[Sec.~3, Table~2]{portisch} illustrate the difficulty. When aligning a loan-management system with a financial data platform, the best automated matcher recovered only about one in eight expert mappings.

Even when a relationship carries over, the evidence needed to act on it can change. Firms in Kumar and Tanniru's study~\citep[Secs.~5--6]{paymentGraph} reused links among vendors, invoices, and payments. A manufacturer may need a goods receipt, while a services firm may need proof of client acceptance. Deployment teams mapped the records and configured local policies. It remains unclear whether an agent can tell when the old evidence falls short and identify the record it now needs.

An agent should recognize which relationships and rules it can safely reuse and where it needs new evidence or expert guidance. The test is whether those judgments yield reliable results on the new task with less manual adaptation.

\subsection{Long-Term Value}\label{sec:5.4}

A financial KG has long-term value when evidence collected for one task helps an agent answer later questions well enough to justify the cost of building and maintaining the graph. The difficult judgment is deciding when to reuse earlier evidence and when to search again. A figure may still be correct for the date it describes but no longer answer a question about the present. Checking every figure again erodes the savings, while unchecked reuse can carry an outdated assumption into later answers.

TrackMem~\citep[Secs.~5.2--5.3 and App.~D.1]{trackmem} reduced how much an agent had to read. In simulated investment conversations, its reader used roughly a quarter to a third of the input tokens required to reread the full history. When TrackMem and similarity search returned the same number of turns, TrackMem answered questions about revised figures more accurately. How many fresh source queries it avoided remains unknown.

TIEM~\citep[Secs.~5.5--5.6 and App.~H.1]{tiem} found that lessons from past events improved stock forecasts overall. In one audit run, the outcomes behind all retrieved lessons were known before each forecast. The harder question is whether the circumstances that made a particular lesson useful still hold for the new event. The study left open which lessons applied and which useful ones the agent missed. Its token counts covered only the final forecast call.

One way forward is to record the conditions under which earlier evidence applies and recheck it when those conditions change. Across a sequence of tasks, researchers could track which evidence the agent reused correctly, overlooked, or misapplied. They could then compare its answers with those from a fresh search on each task. Long-term value depends on whether reuse maintains answer quality and saves enough queries and tokens to justify building and maintaining the graph.

\par\medskip

Across these four challenges, the central question is how far an analyst can rely on a conclusion drawn from a financial KG. Progress would mean that an agent can identify which gaps in the evidence could change an answer, revisit the conclusions affected by a correction, and recognize when a new domain requires different assumptions. Over a sequence of tasks, the practical test is whether those judgments preserve the quality of the analysis while saving enough repeated work to justify building and maintaining the graph.

\bibliographystyle{plainnat}

\bibliography{references}

\end{document}